# Influence of mechanical compliance of the substrate on the morphology of nanoporous gold thin films

Sadi Shahriar[a], Kavya Somayajula[b], Conner Winkeljohn[a], Jeremy Mason[a], Erkin Seker[c,*]

[a]Department of Materials Science and Engineering, University of California - Davis, Davis, CA 95616, USA

[b]Department of Mechanical and Aerospace Engineering, University of California - Davis, Davis, CA 95616, USA

[c]Department of Electrical and Computer Engineering, University of California - Davis, Davis, CA 95616, USA

*Corresponding Author:

Prof. Erkin Seker
3177 Kemper Hall
Department of Electrical and Computer Engineering
University of California, Davis
Davis, CA 95616
USA

Email: eseker@ucdavis.edu
Tel: +1 (530) 752-7300

**Abstract**

Nanoporous gold (np-Au) has found use in applications ranging from catalysis to biosensing where pore morphology plays a critical role in performance. While morphology evolution of bulk np-Au has been widely studied, knowledge about its thin film form is limited. This work hypothesizes that mechanical compliance of the thin film substrate can play a critical role in the morphology evolution. Via experimental and finite-element-analysis approaches, we investigate the morphological variation in np-Au thin films deposited on compliant silicone (PDMS) substrates of a range of thicknesses anchored on rigid glass supports and compare those to the morphology of np-Au deposited on glass. More macroscopic (10s to 100s of microns) cracks and discrete islands form in the np-Au films on PDMS compared to glass. Conversely, uniformly-distributed microscopic (100s of nanometers) cracks form in greater numbers in the np-Au films on glass than on PDMS, with the cracks located within the discrete islands. The np-Au films on glass also show larger ligament and pore sizes possibly due to higher residual stresses compared to the np-Au/PDMS films. The effective elastic modulus of the substrate layers decreases with increasing PDMS thickness, resulting in secondary np-Au morphology effects including a reduction in macroscopic crack-to-crack distance, an increase in microscopic crack coverage, and a widening of the microscopic cracks. However, changes in the ligament/pore widths with PDMS thickness are negligible, allowing for independent optimization for cracking. We expect these results to inform the integration of functional np-Au films on compliant substrates into emerging applications, including flexible electronics.

# 1. Introduction

Nanoporous metals are a subclass of functional nanostructured materials that have drawn significant attention from the research community for a combination of unique characteristics including large surface area-to-volume ratio, high electrical and thermal conductivity, and a network structure of interconnected pores with feature sizes that can be tuned from the nanometer to micrometer range [1–5]. These materials are commonly obtained by a corrosion process called *dealloying* which involves preferential dissolution of one or more less noble elements from an originally homogeneous alloy [6,7]. Nanoporous gold (np-Au) is frequently considered as the prototypical nanoporous metal and is fabricated by the dealloying of AuAg alloys containing 60–80 atomic% Ag, where Ag atoms selectively dissolve while gold atoms diffuse at the surface-electrolyte interface to form a bicontinuous ligament and pore structure consisting mostly of Au atoms [7–11]. The intriguing optical [12], electrical [13–15], and mechanical [16,17] properties of np-Au have created opportunities for applications in a wide range of fields including sensors [18–21], actuators [22–24], catalysis [25], energy storage [26], and biomedical devices [27,28].

The thin film forms of nanoporous metals enable their integration into functional devices (e.g., sensors) via conventional photolithographic techniques [29]. However, thin films supported by an underlying substrate often exhibit residual stresses. These stresses may arise from several sources such as the intrinsic stresses caused by the accumulation of crystallographic defects during film deposition and the thermal stresses from the deposition-induced thermal expansion mismatch between the film and substrate, both of which affect the microstructure and performance of the films [30,31]. Mechanical mismatches, such as compliance differences, between the thin film and the substrate play an important role in residual stress accumulation and relaxation, and ultimately the resulting topographies at scales ranging from nanometers to millimeters. Such mismatches often result in variation in the residual stress-induced channel cracking behavior in the films [32]. For example, if the substrate is much stiffer than the film, the cracks may extend only partially through the film [33], but when the substrate is more compliant than the film, the cracks may extend through the film to the substrate and delaminate along the film-substrate interface [34]. While there are several studies on the deformation and cracking behavior of as-deposited metallic thin films on compliant substrates [35–37], there is limited knowledge on the influence of the substrate compliance on the microstructure of nanoporous metals obtained by dealloying. It was previously demonstrated that the residual stresses in np-Au thin films vary depending on the mechanical constraints imposed on the film (e.g., approximately two times higher residual stress in a substrate-supported blanket film than a microfabricated free-standing film), which leads to different cracking and pore morphologies [38]. Therefore, it is logical to further study the morphology evolution in np-Au thin films in response to variations in substrate compliance and the accompanying residual stress changes. This is particularly important for applications where nanoporous metals could be patterned on compliant substrates such as wearable sensors and flexible electronics [39,40].

In this work, we focus on the crack and ligament-pore morphology evolution in np-Au thin films deposited on compliant silicone substrates of varying thicknesses that are secured to an additional rigid substrate. This system modulates the "effective elastic modulus" of the substrate experienced by the np-Au thin film as a function of the silicone layer thickness. In addition, we provide a comparison of the morphologies in the np-Au films on the compliant substrates to those of the films on the rigid substrate which has a much higher elastic modulus.

## 2. Experimental

### 2.1. Chemicals/materials

Thermo Scientific glass slides (25 × 75 × 1 mm) were used to anchor the polydimethyl siloxane (PDMS) "silicone" substrates. 15 × 8 mm PDMS substrates of thicknesses 0.25, 0.50, 1.59, and 3.18 mm (as per manufacturer datasheet) were prepared from BISCO HT-6240 silicone sheets obtained from Rogers Corporation. Silver (Ag), gold (Au), and chromium (Cr) sputtering targets of 99.95% purity were procured from Kurt J. Lesker. Nitric acid (70%) was purchased from Sigma Aldrich.

*2.2. Sample preparation*

The glass slides were cleaned with isopropanol followed by drying with a nitrogen gun before attaching the PDMS substrates. The glass and PDMS surfaces forming the glass/PDMS interface were exposed to air plasma for 1 minute at 30 W in a PDC-32G plasma cleaner from Harrick plasma. This treatment facilitated the covalent bonding between exposed surfaces of the PDMS and the glass. A Lesker LAB Line sputter system was used to deposit the AuAg alloy film (precursor to np-Au thin film), which consisted of sequential deposition of a Cr adhesion layer (~160 nm-thick), a planar Au intermediate layer (~80 nm-thick), and finally a $Au_{0.24}Ag_{0.76}$ (atomic%) alloy layer (~600 nm-thick). Prior to deposition, the substrate surface was treated in air plasma at 30 W for 2 minutes to improve adhesion of the thin film to the substrate. The AuAg thin film samples were dealloyed for 15 minutes by immersing them in 70% nitric acid heated to 55°C on a hotplate followed by rinsing in deionized water and drying them under nitrogen flow.

*2.3. Post-dealloying characterization*

To characterize the morphology of the np-Au thin films at micro- and nanometer length scales, top-view scanning electron microscope (SEM) images were obtained using a FEI Nova NanoSEM430 microscope at magnifications ranging from 150X to 150,000X. The atomic percentages of Au and Ag in the thin film before and after dealloying were determined by an Oxford X-Max Energy Dispersive X-ray Spectroscopy (EDS) detector in conjunction with a FEI Scios Dualbeam FIB/SEM system. The SEM images were processed and analyzed by a combination of ImageJ [41], GIMP, and MATLAB to quantize the morphological features.

Overlay masks in ImageJ were used on segmented images to obtain pseudo-colored visualizations of the individual islands bound by macroscopic cracks. The thicknesses of the thin films were measured using a Park Systems XE7 atomic force microscope (AFM) over a step of the thin film by masking part of the glass substrate with a Kapton tape during deposition. The AFM was also used to characterize the topographies of the np-Au film surfaces by scanning 50 × 50 µm areas of the film in the tapping mode. The AFM images were processed and analyzed using Gwyddion to extract a "waviness" parameter. Briefly, a total of nine line scans along the x-axis on the AFM topographic images of three different np-Au islands (three lines separated by 10-15 µm per island) were analyzed for each np-Au/substrate combination. The analysis line locations were adjusted to avoid adsorbed particulates that could cause artifacts in the scan profile. The waviness profiles along those lines were extracted using a cut-off wavelength of 4 µm to filter out the roughness. The average waviness for each line scan (i.e., the average heights of the waviness profile along each line scan) were then computed.

*2.4. Simulations*

To corroborate the results of the experiments, finite element analysis (FEA) simulations were performed using COMSOL Multiphysics software. These involved simulating np-Au thin films on a PDMS substrate with the PDMS fixed at the bottom to mimic the rigid glass substrate. The first set of simulations was performed to calculate the "effective elastic modulus" of the anchored PDMS substrates wherein the PDMS layer was 5 mm in both length and width with the thickness varying from 0.01 mm to 5 mm. A tetrahedral mesh with an element size range of 0.501 µm to 2.5 µm was used for this set of simulations.

The second set of simulations was performed to estimate the elastic strain energy in the thin film-substrate system, and the horizontal and vertical edge displacements at the metal film/substrate interface before and after dealloying. We refer to the Cr and Au layers as the *adhesion layers* for the rest of the paper. The PDMS layer was 2.5 mm in both length and width for these simulations with the thickness varying from 0.05 mm to 5 mm. The thicknesses of the AuAg (post-deposition) and np-Au (post-dealloying) films were 600 nm and 500 nm, respectively, and the Cr and Au adhesion layers had thicknesses of 160 nm and 80 nm, respectively. A free quadrilateral mesh with an element size range of 0.079 µm to 318 µm was used. A swept mesh was applied for all the layers (Figure SI 11), with the number of elements through the thicknesses of AuAg and np-Au films, adhesion layers, and PDMS being 8, 10 and 5 respectively.

*2.5. Statistical Analysis*

A minimum of two different samples with a minimum of three different images per sample per length-scale was used for statistical comparisons. A student's t-test was performed to compare two different sample groups, with *p*-values below 0.05 deemed statistically significant. The statistical tests were performed with OriginPro.

3. **Results**

Figure 1 shows the morphologies of the precursor AuAg films on glass and PDMS substrates at two different magnifications. At low magnification (Figure 1a), it is apparent that cracks have initiated in the films on the PDMS substrates but not on the glass substrate. However, no cracks are visible in any of the films at the length scale of grains (Figure 1b).

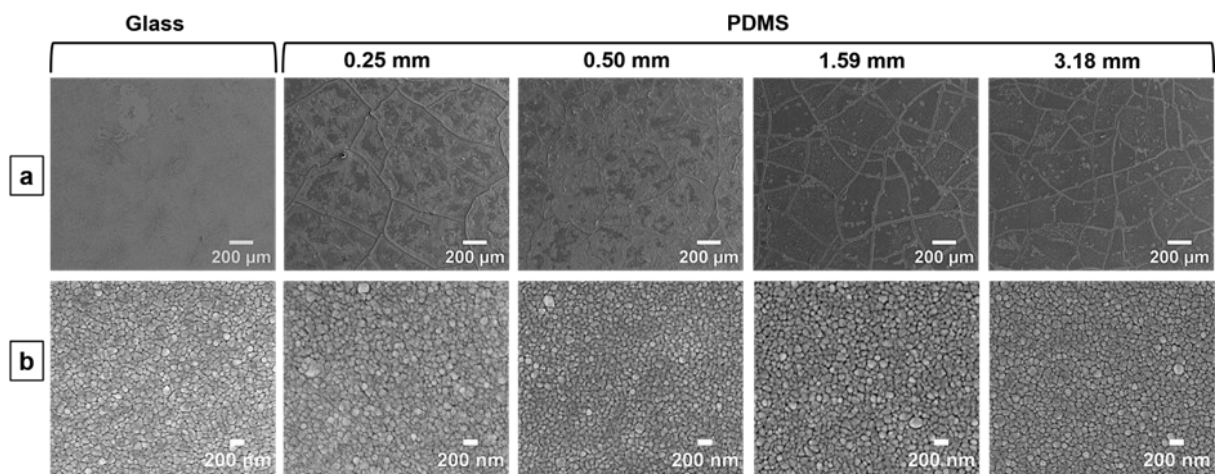

**Figure 1**. Top-view SEM images of as-deposited AuAg precursor thin films on glass and PDMS of varying thicknesses at (a) low (150X), and (b) high (50kX) magnifications.

Figure 2 shows the three different types of morphological features in the np-Au thin films at different length scales: macroscopic cracks, microscopic cracks, and ligaments and pores.

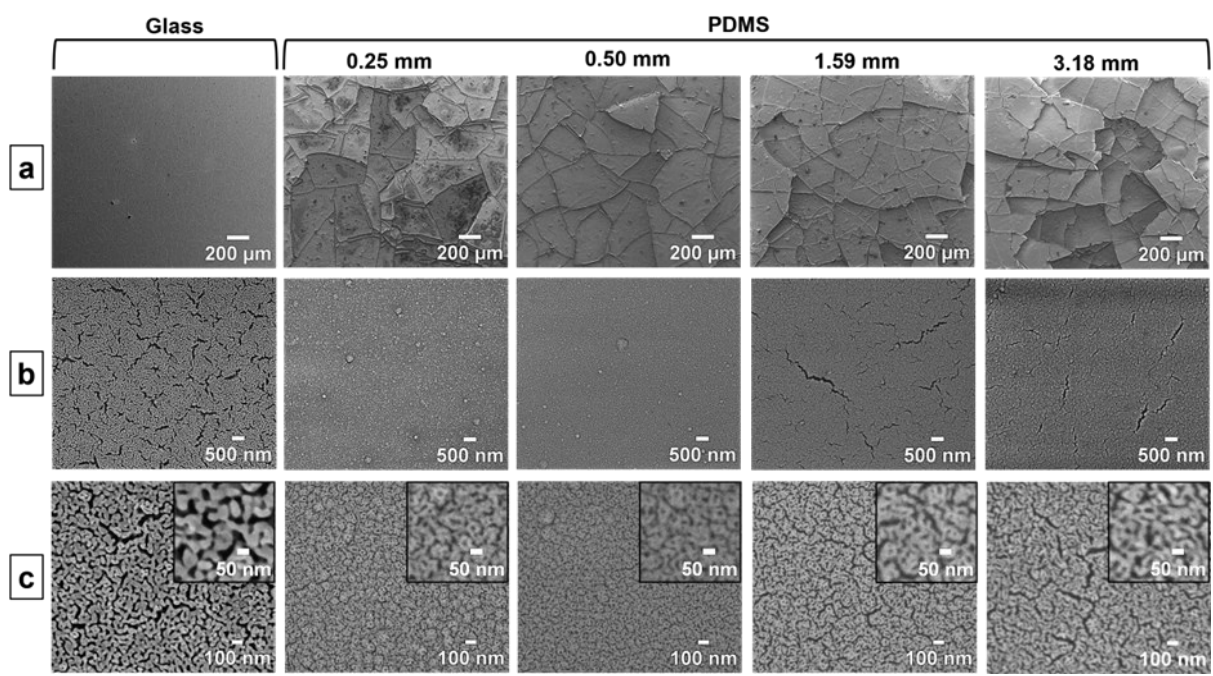

**Figure 2**. Top-view SEM images of (a) macroscopic cracks at 150X, (b) microscopic cracks at 35kX, and (c) ligaments and pores at 150kX magnifications in the np-Au thin films on glass and PDMS of varying thickness.

Here, we investigate the variation in these features across np-Au films deposited on compliant PDMS substrates (np-Au/PDMS) of thicknesses 0.25, 0.50, 1.59, and 3.18 mm that are covalently bonded onto glass slides of 1 mm thickness. In addition, the morphology of the np-Au film on "infinitely" rigid glass (np-Au/glass) is compared to that of np-Au/PDMS. As evident in Figure 1, the precursors to macroscopic cracks in the np-Au/PDMS films are present following the deposition step, but the microscopic cracks appear after dealloying (Figure 2). The films on glass are instead free from macroscopic or microscopic cracks prior to dealloying but show profuse microscopic cracking after dealloying.

*3.1. Macroscopic cracks*

As seen in Figure 2a, the macroscopic cracks are present throughout the film and are visible at low magnification (150X) SEM images; the segmentation of the macroscopic cracks in np-Au is shown in Figure SI 4. On the PDMS substrates, individual islands are separated from each other by macroscopic cracks. However, the cracks in the film on glass, although ubiquitous at this length scale, are smaller than those in the films on PDMS and do not form enclosed regions as on PDMS. Figure 3 shows how the macroscopic crack-to-crack distances vary among different np-Au film/substrate combinations. The crack-to-crack distances for np-Au/PDMS were taken to be the island widths and were quantified by first segmenting the SEM images and then measuring the major and minor axes of the elliptical outlines of the segmented regions (details in Section 1.1 of the SI). Since the cracks on np-Au/glass do not form enclosed regions, the elliptical outline method is not suitable for measuring the inter-crack distance. We instead used a custom MATLAB code [14] to scan horizontally and vertically along the segmented image and find inter-crack distances along these directions. The values computed with this approach were scaled up by 1.53 to match the results provided by the elliptical outline method (rationale described in Section 1.4 of the SI). The areas of individual islands in the np-Au/PDMS films get smaller, the number of islands increases, and the distance between the cracks decreases with increasing PDMS thickness. The increase in crack density (number of cracks per unit area) with PDMS thickness is clearly visible from the segmented images in the inset of Figure 3, and is further evidenced by the average crack-to-crack distance decreasing from 260 ± 7 μm to 167 ± 3 μm (the errors are the standard error of the mean) for PDMS thicknesses from 0.25 mm to 3.18 mm. In addition, the distribution in the distance values becomes narrower for thicker PDMS substrates (1.59, 3.18 mm) than those for the lower thickness (0.25, 0.50 mm). The average crack-to-crack distance for the np-Au/glass is 117 ± 0.53 µm and is substantially smaller than that for even the np-Au/PDMS film on 3.18 mm-thick PDMS.

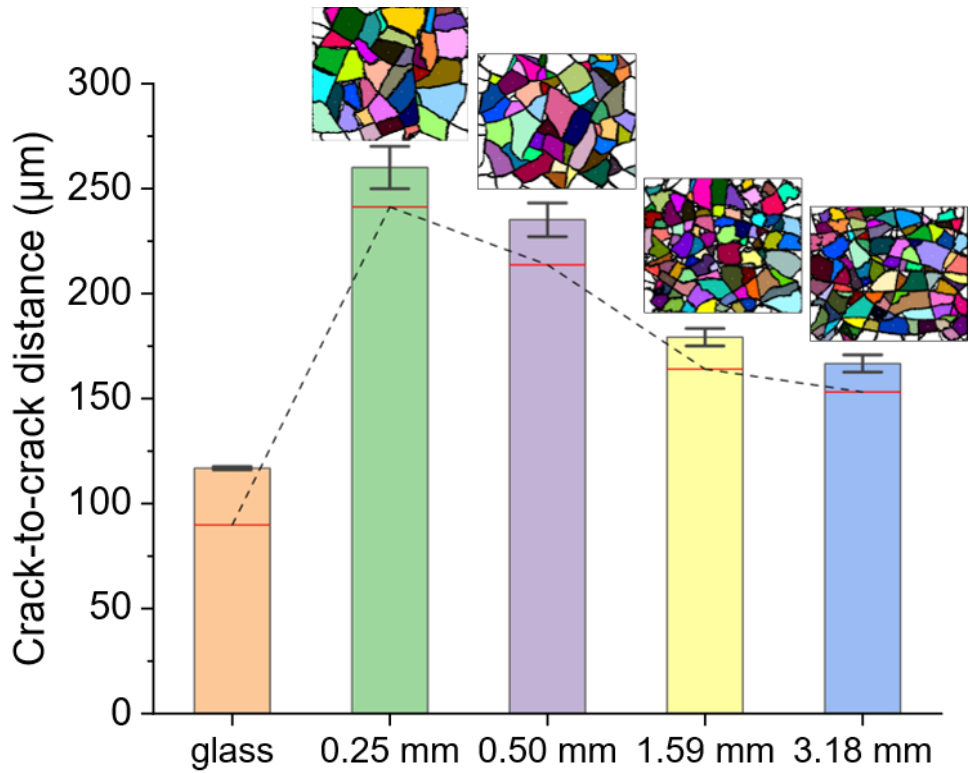

**Figure 3**. The average crack-to-crack distance for the macroscopic cracks observed in np-Au/glass and np-Au/PDMS of varying substrate thickness. The islands bound by the cracks in np-Au/PDMS are shown in the inset with pseudo-color segmentation. The red lines inside each bar denote the median crack-to-crack distance and the dashed line goes through the median values. The error bars denote the standard error of the mean.

## 3.2. Microscopic cracks

Higher magnification SEM (35kX) reveals the presence of "microscopic" cracks in the thin films as shown in Figure 2b. To quantify the microscopic cracking, the SEM images were segmented to distinguish the cracks as black pixels against a white background (Section 1.2 of the SI). Microscopic cracks are more prevalent in np-Au/glass films compared to np-Au/PDMS films; the percent crack coverage (Figure 4) for np-Au/glass is 7.6% which is more than twice the maximum crack coverage observed for np-Au/PDMS films. The inset of Figure 4 confirms that np-Au/glass films have a significantly larger population of microscopic cracks than the average population in the np-Au films on any thicknesses of PDMS.

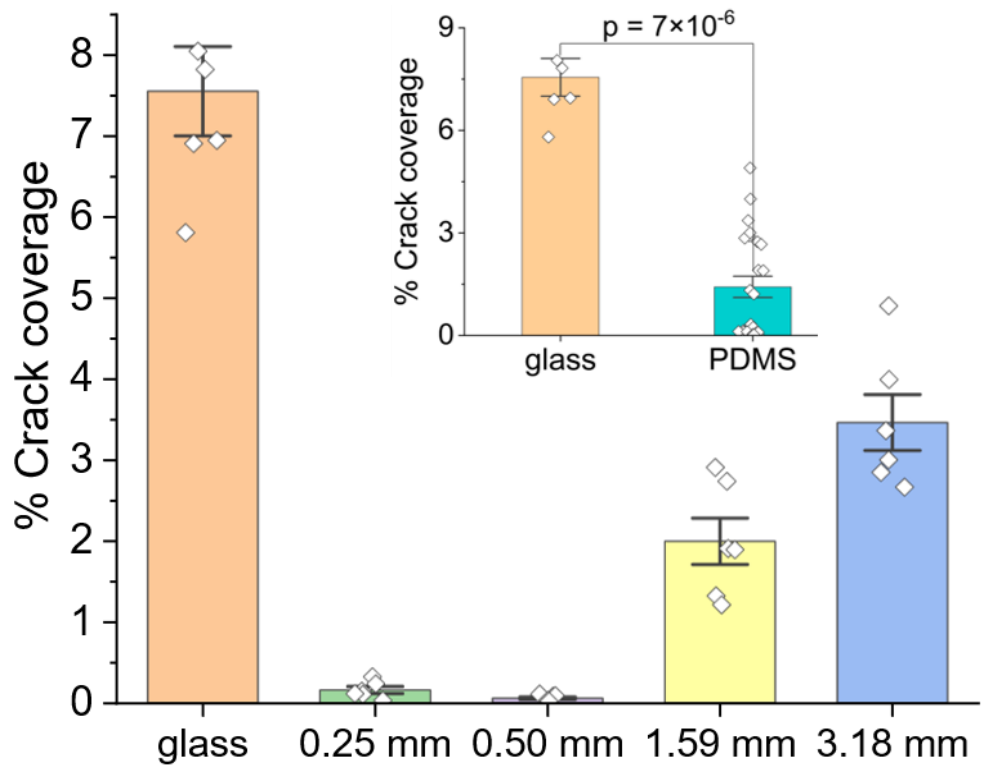

**Figure 4**. Variation in the percentage of the crack coverage in np-Au/glass and np-Au/PDMS films with varying substrate thickness. The inset shows the comparison of the crack coverage between np-Au/glass and the average for np-Au/PDMS films, where a p-value less than 0.05 denotes a statistically significant difference. The error bars represent the standard error of mean.

Interestingly, np-Au/PDMS films on the 0.25 and 0.50 mm-thick PDMS substrates display negligible microscopic cracking whereas cracks appear in np-Au films on thicker PDMS substrates, gradually increasing to a percent crack coverage of 3.5% on 3.18 mm-thick PDMS. That is, the maximum percent crack coverage on thicker PDMS is ~25 times higher than the maximum coverage on thinner PDMS.

## 3.3. Ligaments and pores

At the smallest length scale, we investigated the ligaments and pores at 150kX magnification in the np-Au thin films as shown Figure 2c. The ligament and pore widths were measured by segmenting the SEM images and applying a custom MATLAB script to analyze the segmented images (details in Section 1.3 of the SI). The width distribution of the ligament and pores are presented as violin plots to capture the distribution of the ligament/pore sizes (Figure 5).

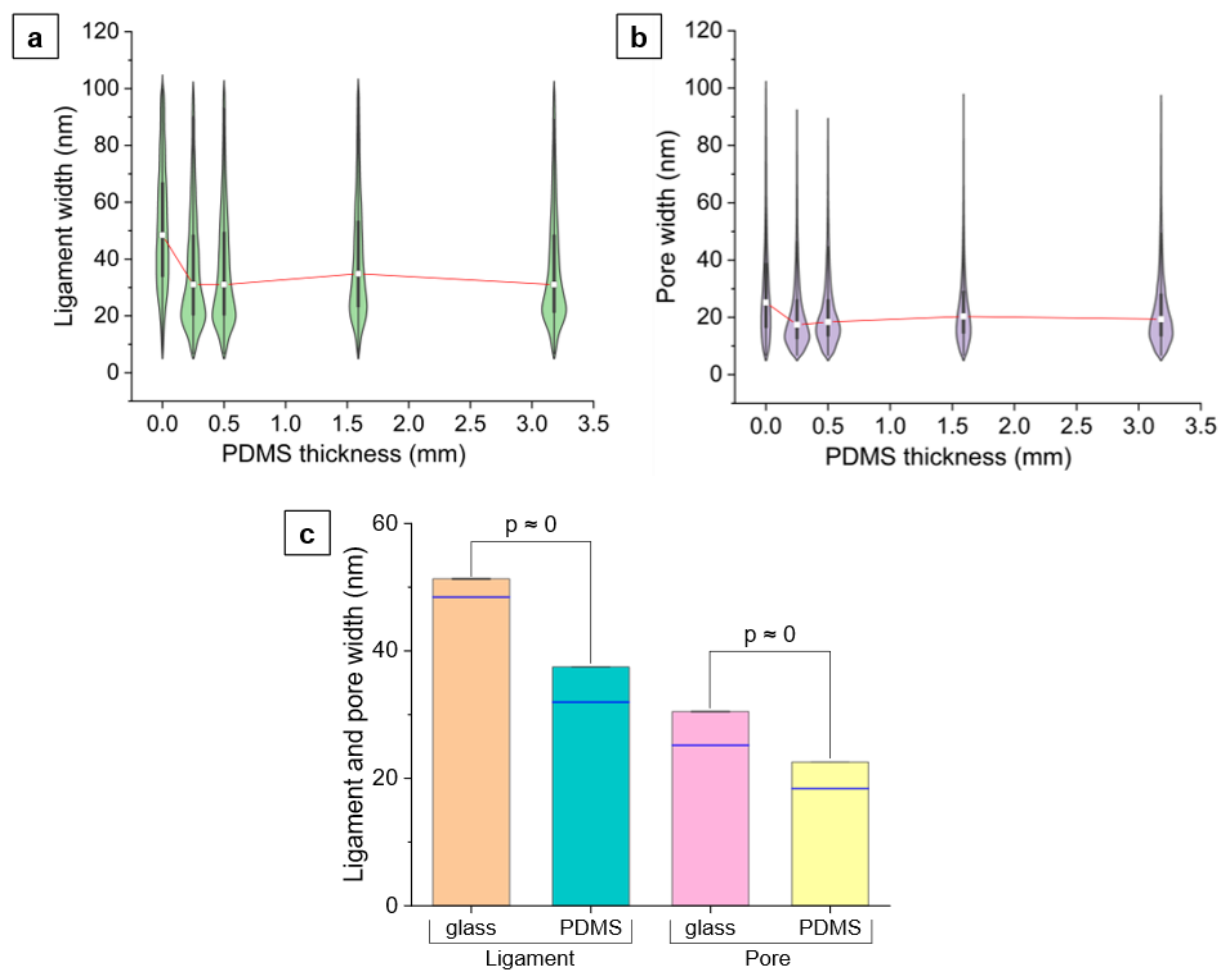

**Figure 5**. (a, b) Ligament width (a) and pore width (b) distribution in np-Au/glass (0 mm PDMS thickness) and np-Au/PDMS of varying substrate thickness shown by violin plots. The boxes inside the violins range from the first to the third quartiles, the whisker lengths show 1.5 times the interquartile range, and the white squares denote the median values. The median ligament and pore width on glass are larger than those on PDMS, but they do not display a marked variation with varying PDMS thickness. The red trendlines through the medians are visual guides only. (c) Comparison of the average ligament and pore widths of np-Au/glass to those of np-Au/PDMS averaged over all the PDMS thicknesses. A p-value less than 0.05 indicates a statistically significant difference. The p values, being very small numbers in this case, have been approximated as zero. The blue lines correspond to the median values and the error bars denote standard error of the mean (negligible due to very small values).

The median ligament and pore widths in np-Au/glass are 48 and 25 nm, respectively, which are larger than the maximum median values of 35 nm and 20 nm for the ligament and pore widths for np-Au on 1.59 mm-thick PDMS. That is, the ligament width has greater absolute and relative changes than the pore width when going from a stiff to a compliant substrate. The widths of the ligaments on glass also show a wider distribution (Figure 5). However, neither the ligament nor the pore width in np-Au/PDMS films show a marked variation with substrate thickness, with the median ligament widths ranging from 31 to 35 nm and the median pore widths from 17 to 20 nm. As shown in Figure 5c, the average ligament and pore widths for np-Au/glass are significantly larger than those for np-Au/PDMS averaged over all the PDMS thicknesses.

### 3.4. Finite Element Simulations

To simulate the effective elastic moduli of anchored PDMS of different thicknesses $h_P$, a two-dimensional PDMS mesh consisting of tetrahedral elements with a minimum size of 0.501 μm was generated (Section 2.4). A force of $f = 0.0001$ N along the x direction was applied to two points that were $w_0 = 200$ μm apart on top of the PDMS, and the resulting distance $w$ between the points was recorded (Figure 6a). The effective elastic modulus was defined by equating the average strain energy density in the PDMS with that of a homogeneously strained linear elastic isotropic solid in the following way. The work per depth of material is $f(w - w_0)$, and the PDMS volume per depth of material between the points of contact is $h_P w_0$, giving a strain energy density of $f\varepsilon/h_P$ where $\varepsilon = (w - w_0)/w_0$ is the linear strain on the PDMS surface. Equating this with the elastic strain energy $E_P \varepsilon^2 / 2$ in an isotropic linear elastic solid and solving for the Young's modulus $E_P$ as a function of PDMS thickness $h_P$ gives

$$E_P = \frac{2f}{h_P \varepsilon} \tag{1}$$

Figure 6b shows the variation in the effective elastic modulus of the anchored PDMS substrates for nine different thicknesses (0.001 mm to 5 mm) as defined by Equation 1, and the modulus decreases monotonically with increasing PDMS thickness. Note that the modulus at the lowest PDMS thickness (0.01 mm) far exceeds the physical modulus of glass (~70 GPa [42]) by virtue of Equation 1 since the simulations and the equation modeled the PDMS as being anchored to an infinitely rigid material. Experimentally, the maximum effective modulus of PDMS should not exceed 70 GPa and should only reach this value when the PDMS thickness approaches zero.

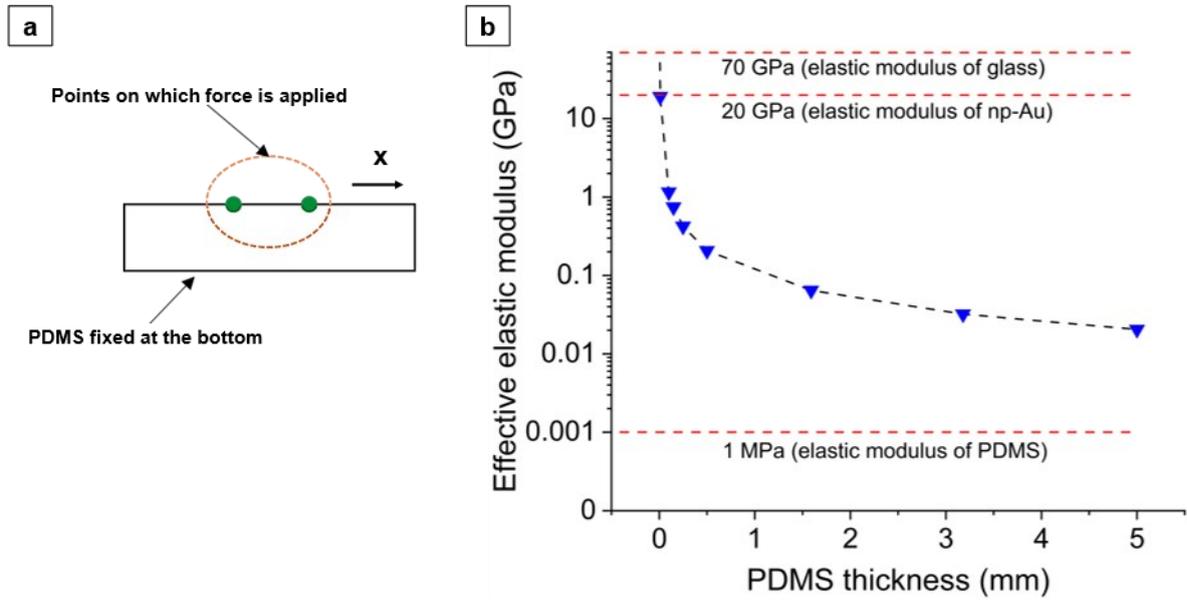

**Figure 6**. (a) Schematic showing the simulation setup to compute the effective elastic modulus of PDMS at the surface. (b) Effective elastic modulus of the PDMS substrate at the free surface as a function of the PDMS thickness. The black dashed line is a visual guide only.

Figure 7 shows the elastic strain energies in the top metal layers and anchored PDMS substrates of ten different thicknesses (0.05 mm to 5 mm) before and after dealloying as obtained from the simulation setup described in Section 2.4. Figure 7a also includes the strain energies in AuAg and np-Au on glass. A thermal strain corresponding to a stress of 100 MPa was applied to the metal layers to calculate the post-deposition strain energies. This strain was then used as a pre-strain in the metal layers for the dealloying simulations. AuAg on glass has the highest strain energy of $6.3 \times 10^{-7}$ J, with this dropping to $1.5 \times 10^{-7}$ J in np-Au after dealloying (Figure 7a). The strain energy in the AuAg film on 0.05 mm-thick PDMS drops from $2.9 \times 10^{-8}$ J to $1.6 \times 10^{-8}$ J in the np-Au film after dealloying, with both of these values decreasing with increasing PDMS thickness (Figure 7a and 7b). The strain energy for the PDMS in the post-deposition state similarly decreases with PDMS thickness from a maximum of $2.3 \times 10^{-8}$ J for the 0.05 mm-thick PDMS to $2.1 \times 10^{-9}$ J for the 5 mm-thick PDMS (Figure 7b). The strain energy of the total film-PDMS system consistently drops by a factor of ~1.3 after dealloying as shown in

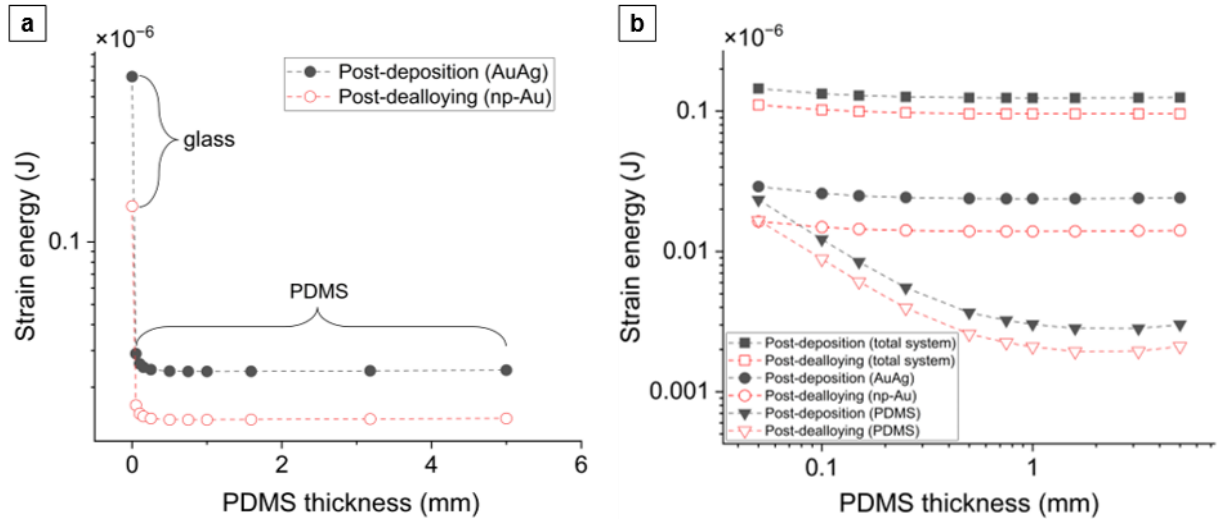

**Figure 7**. Post-deposition and post-dealloying variation in the elastic strain energy in (a) the AuAg and np-Au layers on glass and anchored PDMS of different thicknesses, and (b) the total film-substrate system, the AuAg and np-Au films, and the anchored PDMS substrates of varying thicknesses. The y axis in (a) and both x and y axes in (b) are in log scale. The dashed lines are visual guides only.

The horizontal and vertical deformation at the film-substrate interface obtained by the simulation is illustrated in Figure 8. As shown in Figure 8a, the post-dealloying deformation at the np-Au/glass interface is very small as expected. Conversely, the deformation at the np-Au/PDMS interface is significantly larger with in-plane (horizontal) compressive deformation and out-of-plane (vertical) deformation. The post-deposition average horizontal edge displacement at the np-Au/PDMS interface increases from 1.07 μm for 0.05 mm-thick PDMS to 1.12 μm to 5 mm-thick PDMS (Figure 8b) and the average vertical edge displacement increases from 0.69 μm for 0.05 mm-thick PDMS to 0.94 μm for 3.18 mm-thick PDMS (Figure 8c). In addition, the post-deposition average horizontal and vertical displacements at the AuAg/PDMS interface are slightly larger than the post-dealloying displacements at all PDMS thicknesses (Figure 8b and 8c).

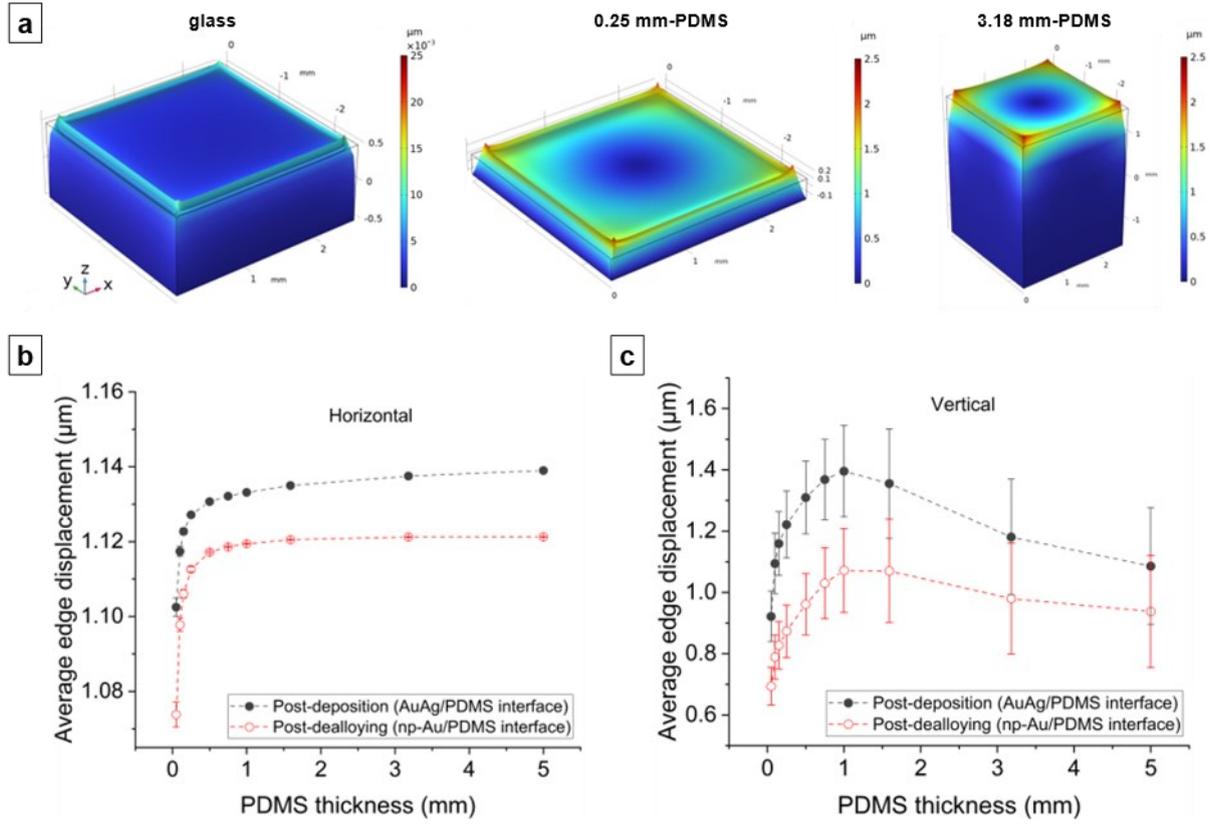

**Figure 8.** (a) Simulated deformations for post-dealloying states with representative substrate types of glass, 0.25 mm-thick PDMS and 3.18 mm-thick PDMS. Average (b) horizontal and (c) vertical post-deposition and post-dealloying edge displacements at the AuAg/PDMS and np-Au/PDMS interfaces as a function of PDMS thickness. The error bars denote standard deviation. The dashed lines are visual guides only.

## 4. Discussion

The most prominent differences in the np-Au film morphologies occur between the np-Au/glass and np-Au/PDMS/glass systems and constitute the main experimental observation. This is attributed to the difference in the surface stiffness experienced by the metal thin film deposited on rigid glass versus compliant PDMS with elastic moduli of ~70 GPa [42] and ~1 MPa [43] respectively. Secondarily, since the PDMS substrate is anchored onto a glass slide, the thickness of the PDMS also affects the effective elastic modulus of the substrate as shown in Figure 6. The elastic mismatch between the film and substrate is quantified by the Dundurs' parameters, α and β [34,44,45], which are defined as [44]:

$$\alpha = \frac{\bar{E}_f - \bar{E}_s}{\bar{E}_f + \bar{E}_s}, \qquad \beta = \frac{\bar{E}_f f(\nu_s) - \bar{E}_s f(\nu_f)}{\bar{E}_f + \bar{E}_s} \qquad (2)$$

where the subscripts $f$ and $s$ correspond to the film and substrate, respectively, $\bar{E} = E/(1 - \nu^2)$ and $f(\nu) = (1 - 2\nu)/[2(1 - \nu)]$ for the plane strain condition, $E$ is the elastic modulus, and $\nu$ is Poisson's ratio. Equation (2) indicates that $\alpha > 0$ for a stiffer film on a more compliant substrate with $E_f > E_s$, as for np-Au/PDMS, whereas $\alpha < 0$ for np-Au/glass. This sign reversal in the Dundurs' parameters when switching from a compliant to a stiff substrate should change the dominant mode of morphological feature evolution, consistent with the results of this work. Here, we will separate the discussions into the *main effects* (comparison of morphological features between glass and PDMS) and the *secondary effects* (comparison of morphological features within different PDMS thicknesses).

*4.1. Main effects*

*4.1.1. More macroscopic cracking in np-Au/PDMS compared to np-Au/glass*

As seen in Figure 1, macroscopic cracks start forming in the films on PDMS after sputter deposition and cracking becomes more significant after dealloying, whereas post-deposition cracks are absent in the films on glass. Tensile residual stresses up to 110 MPa were reported in 40 nm-thick sputter-deposited Ag films [46]. As the tensile strengths of Au and Ag are 100 and 140 MPa, respectively [47], the tensile strength of our alloy film containing 24 at% Au and 76 at% Ag is estimated to be ~130 MPa with the rule of mixtures [48]. The residual stress in the as-deposited films possibly exceeds this value as evidenced by the formation of cracks in the films on PDMS to partially relieve the residual stress. Intergranular cracking in 100 nm-thick free-standing Au thin films under tensile stress was previously reported where the cracks extend along multiple grain boundaries (GBs) through GB sliding and shearing [49]. The mechanism of crack formation and extension in the as-deposited AuAg films on PDMS are likely similar where the residual tensile stress drives the cracks along the GBs. However, this process is aided by deformation of the compliant PDMS substrate at the PDMS-metal interface ultimately resulting in discrete islands bound by the cracks. On the other hand, the absence of cracks in the as-deposited films on glass can be explained by the high elastic modulus of glass (~70 GPa), effectively leading to a zero-displacement boundary condition where there is residual tensile stress in the metal film without cracking. The residual stress in the as-deposited AuAg films on stiff silicon wafers were measured to be ~100 MPa in a previous study [50]. Our AuAg thin film deposited using the same procedure should result in a comparable residual stress for AuAg.

Considerable volume contraction in the np-Au film during dealloying [51] results in tensile stresses which are partially relieved by the formation of cracks [38,52]. A drastic drop of the residual stress from ~100 MPa in the as-deposited AuAg film to only ~20 MPa in the dealloyed np-Au film has been reported [50]. The dealloying-induced stresses together with the post-deposition stress result in crack formation in the np-Au/glass films at different length scales. However, the macroscopic cracks in np-Au/glass occur at a smaller length scale than those on np-Au/PDMS and the crack patterns are different. In the np-Au/PDMS films, the pre-existing macroscopic cracks in the precursor film widen and additional macroscopic cracks appear due to dealloying stresses, resulting in the formation of discrete islands which are not observed in np-Au/glass films.

*4.1.2. More microscopic cracks in np-Au/glass compared to np-Au/PDMS*

In contrast to the substrate-dependent trend of macroscopic cracks described in the previous section, np-Au/glass films exhibit a higher crack surface coverage compared to the np-Au/PDMS films (Figure 4). The microscopic cracks in the np-Au/glass are distributed uniformly throughout the film, whereas the cracks in np-Au/PDMS films show a hierarchical pattern with the microscopic cracks forming inside the discrete islands, predominantly on the thicker PDMS substrates (discussed further under *Secondary Effects*).

The mechanisms for microcrack formation in the np-Au films differ for glass and PDMS

substrates. Initially the thin films in the stack (AuAg, Au, Cr) have similar effective elastic moduli after deposition. After dealloying, the elastic modulus of the top film layer (np-Au) is reduced by approximately four times. The eigenstrain that develops in the np-Au film during dealloying increases the tensile stress in the film overall despite the increased compliance of the np-Au, while the compatibility condition at the substrate interface subjects the substrate to a compressive stress as evidenced by the increasing deformations in PDMS as a function of its thickness (Figure 8). However, the high elastic modulus of the glass means that the glass substrate does not significantly deform to accommodate the eigenstrains in the np-Au film, causing the residual stress in the np-Au/glass films to surpass the film's tensile strength and leading to crack formation to release the strain energy (Figure 2b). Conversely, the much lower effective elastic modulus of PDMS (Figure 6b) allows for the compression of PDMS in the plane of film, leading to out-of-plane buckling due to the substantial Poisson effect (observed as large vertical deformations in the simulation results in Figure 8c). It is important to note that the simulations do not directly capture the buckling observed in experiments due to the idealized defect-free substrate and metal stack structures. For the actual experimental conditions, imperfections in the layers initiate buckling (as shown in Figure 10 below). The buckling can be a lower-energy deformation mode than crack formation and mitigates the microscopic cracks at lower PDMS thicknesses (Figure 2b). However, the increasing out-of-plane buckling magnitude with PDMS thickness plausibly results in high bending stresses at the convex regions (buckling maxima), leading to the emergence of microscopic cracks in np-Au on thicker PDMS substrates (discussed further in Section 4.2).

*4.1.3. Larger ligament and pore sizes in np-Au/glass compared to np-Au/PDMS*

As shown in Figure 5, the ligament widths in np-Au/glass are significantly larger than those in np-Au/PDMS. It was previously observed that np-AuPd films obtained by dealloying precursor alloys (AuPdAg) deposited on curved polyimide substrates displayed coarser ligaments and less residual silver at convex regions of the substrate [53]. This was attributed to higher local stresses at the convex regions, causing silver to dissolve faster during dealloying, which exposes the Au and Pd atoms to nitric acid for longer duration. Increased diffusivity of surface atoms (Au and Pd) coarsens the ligaments [53]. As a corollary to this observation, we hypothesize that the higher tensile stress in thin films on rigid glass compared to compliant PDMS should have a similar effect. We measured the residual silver (at%) in np-Au on glass and on PDMS (0.50 and 1.59 mm-thick PDMS) using EDS. In agreement with our hypothesis, the residual Ag in np-Au was ~4% lower for glass compared to PDMS (Figure 9a) and there was no statistically-significant difference between the two extreme PDMS thicknesses (Figure 9b). This suggests that higher tensile stress in np-Au on glass compared to the PDMS may be playing a role in the larger ligament thickness for np-Au on the glass substrate.

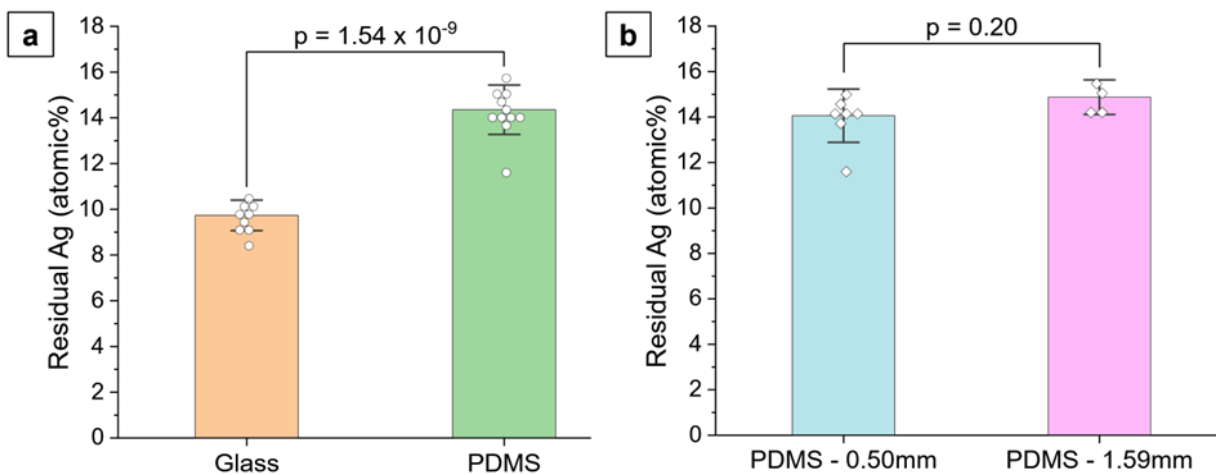

**Figure 9**. Comparison of the average residual silver content after dealloying between (a) np-Au/glass and np-Au/PDMS, and (b) np-Au/0.50 mm-thick PDMS and np-Au/1.59 mm-thick PDMS. A p-value of less than 0.05 corresponds to statistically distinct groups. The error-bars show the standard deviations.

*4.2. Secondary effects*

As discussed in the previous section, there were significant differences in thin film morphology between glass and PDMS substrates where the latter has at least two orders of magnitude lower effective elastic modulus. Although not as prominent, there were also morphological differences in np-Au films on PDMS substrates as a function of the substrate thickness, and these are referred to as *secondary effects*. With increasing PDMS thickness, these effects include a marginal decrease in the island widths indicating a higher density of macroscopic cracks (Figures 2a and 3), an increasing number of microscopic cracks (Figure 2b and 4), and increasing microscopic crack widths (Figure SI 8 and 9). In general, these morphological changes are attributed to the decreasing effective elastic modulus of the PDMS substrate with increasing substrate thickness (Figure 6b). We will focus the discussion on the emergence of microscopic cracks in np-Au, since the largest morphological changes are observed at this length scale.

As mentioned in Section 4.1.2, we attribute the emergence of microcracks in np-Au films on thicker PDMS substrates to increasing buckling due to their decreasing effective moduli (Figure 6b). We hypothesize that np-Au film topography should exhibit larger out-of-plane features as a function of increasing PDMS thickness, reminiscent of the larger buckling amplitudes in precursor AuAg films. We used AFM to characterize topographies of the np-Au films on glass and on PDMS of different thicknesses (Figure 10a). As expected, the rigid glass surface does not exhibit any buckling features. In contrast, the compliant PDMS surfaces display buckling-related features with higher out-of-plane magnitude with increasing PDMS thickness. For the thinner PDMS substrates (0.25 mm, 0.50 mm), the buckling amplitude is smaller, and the waveform is smoother (quasi-sinusoidal with less abrupt changes). The out-of-plane magnitudes were reported as a "waviness" parameter which quantifies the longer spatial wavelength component of the surface topography and is obtained by filtering out the shorter wavelength component (roughness) using a cut-off wavelength of 4 μm [54,55]. The lack of buckling on np-Au/glass is evident by the very low average waviness (Figure 10b). The transition from glass to PDMS results in a sharp rise in the average waviness which increases with increasing PDMS thickness because of the transition from the smoother to sharper waveforms but approaches a plateau for the thicker PDMS as shown in Figure 10b. These topographical features are attributed to the initial PDMS buckling following deposition-related residual stresses. Upon dealloying, the collective elastic modulus of the metal stack decreases (since np-Au's elastic modulus is around four times lower than that of precursor AuAg). While this relieves the strain energy in the system (Figure 7) and reduces the compressive deformation (Figure 8), the brittle nature of np-Au at the macro-scale [56] results in microcracks likely at buckle peaks due to tensile bending stresses (Figure 2 and 4). The pre-/post-dealloying buckling in the substrate-supported thin films here has similarities to doubly-clamped free-standing AuAg beams with various buckling amplitudes and the resulting np-Au beams with corresponding residual stresses, reported previously [57]. In that study, prescribed buckling of AuAg beams (hence compressive pre-strain due to buckling) compensates tensile stress accumulation during dealloying, observed as a reduced occurrence of tensile fracture of np-Au beams. Similarly, PDMS surface buckling is expected to result in reduced cracking in np-Au films on PDMS compared to those on glass (Figures 2b and 4) for thinner PDMS substrates with smaller buckling amplitudes that results lower localized tensile stresses than the fracture strength of np-Au. For thicker PDMS substrates with larger edge deformations (hence larger expected buckling amplitudes), the localized tensile stress at the buckle peaks should be responsible for the microcracks. Finally, the residual buckling in np-Au on 0.50 mm-thick PDMS (Figure 10a) suggests that the compressive strain in PDMS (and the resulting buckling) is partially relieved by the reduction in the np-Au elastic modulus upon dealloying (also observed as reduced strain energy in simulations, Figure 7b).

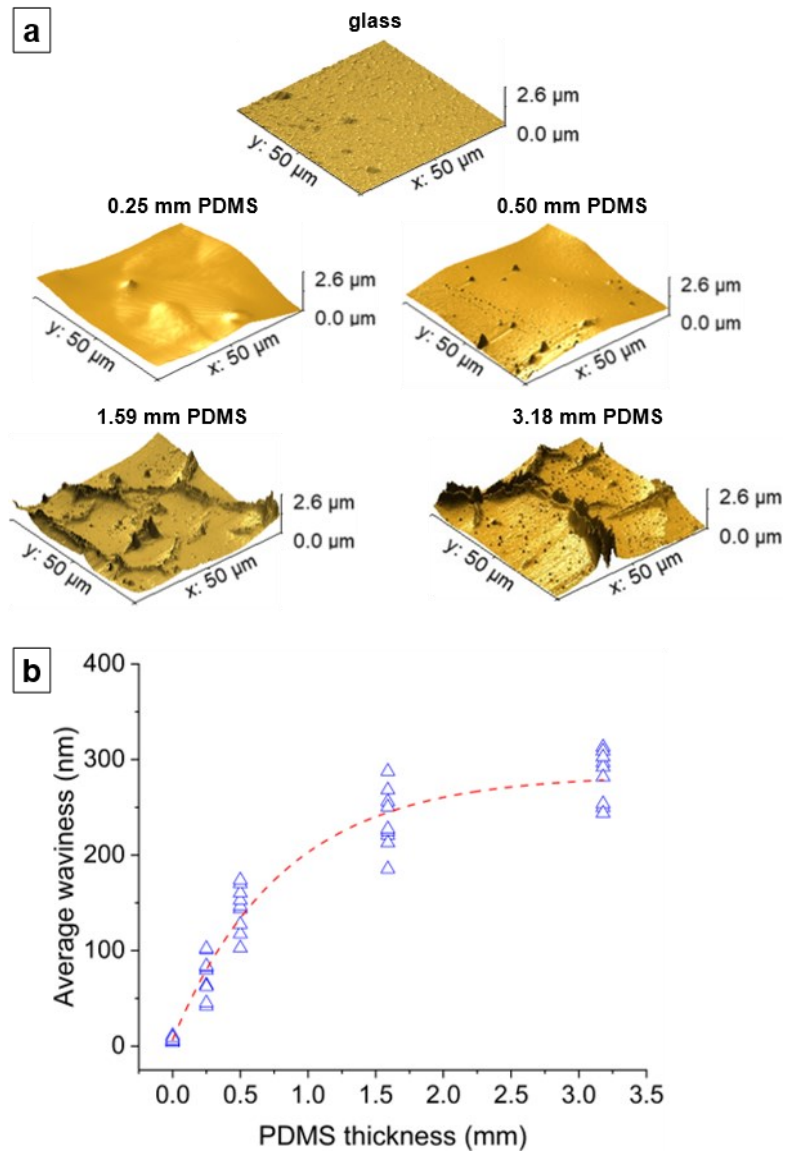

**Figure 10**. (a) AFM topographies of the np-Au film surface on glass and PDMS of different thicknesses showing absence of buckling in np-Au/glass and variation of buckling in np-Au/PDMS. (b) Average waviness values of the surface topography of np-Au films as a function of PDMS thickness (zero thickness denotes the glass substrate) obtained by multiple line scans of different np-Au islands.

As the PDMS thickness (hence effective surface compliance) increases further, the buckling amplitude is expected to increase and possibly exhibit higher order buckling modes, both of which would result in higher local stresses in convex regions. When the buckling amplitude exceeds a critical value, the tensile stress in the np-Au causes the film to rupture at the peaks of the buckles, as shown by others [37]. These ruptures would appear as the large, abrupt and non-periodic topographical features in np-Au films on thicker PDMS substrates (Figure 10a). Taken together, larger buckling amplitudes results in larger crack widths as a function of increasing PDMS thickness (Figure SI 9).

It is important to note that with decreasing PDMS thickness, the film cracking behavior should gradually approach that of glass. This likely occurs around a thickness of 0.01 mm at which the effective elastic modulus passes above the modulus of np-Au (showed by the dashed red line in at 20 GPa in Figure 6) resulting in the sign reversal of the Dundurs' parameter α from positive to negative (Equation 2) and the behavior of a thin film on a stiff substrate. The np-Au films on the thinnest (0.25, 0.50 mm) PDMS substrates in our experiments do not show the crack pattern seen in np-Au/glass films since the PDMS for these thicknesses is still two orders of magnitude more compliant compared to glass and hence should not be expected to display np-Au/glass behavior. Figure 11 qualitatively summarizes the proposed mechanisms of cracking and topographical features among the film-substrate combinations used in this study.

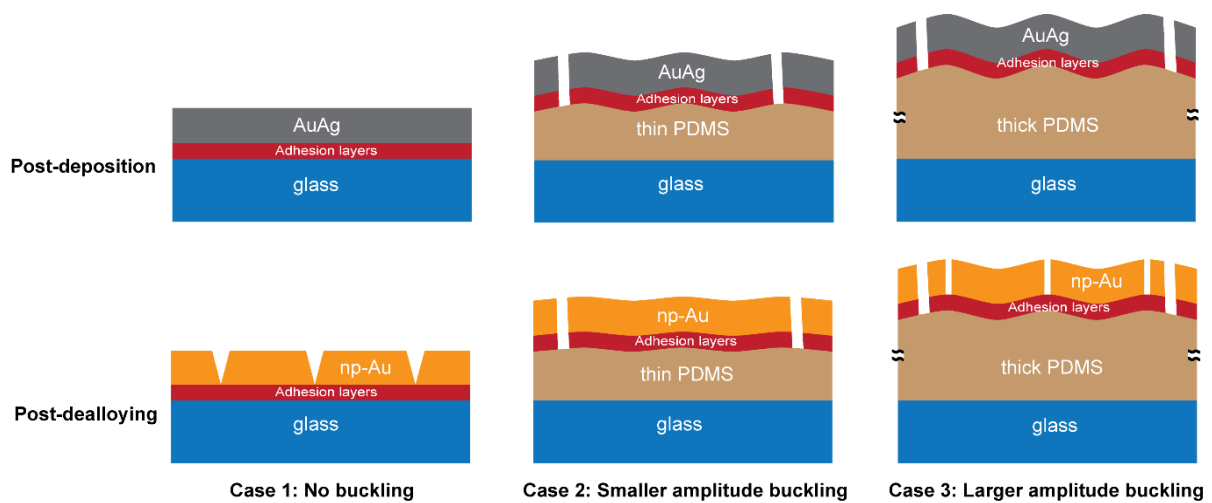

**Figure 11**. A schematic showing the proposed mechanisms of crack formation in np-Au on rigid glass and compliant PDMS substrates. On glass (Case 1), there is no buckling at the glass surface due to the high elastic modulus of glass leading to wedge-shaped anchored cracks with smaller openings. On PDMS (Case 2 and 3), the crack widths in np-Au are larger due to the lower effective modulus of PDMS. For the thinner PDMS (Case 2), the buckling amplitudes are smaller in the post-deposition and post-dealloyed films and are not sufficient to lead to large enough tensile stresses at the buckle peaks to cause rupture and microscopic cracks. For the thicker PDMS (Case 3), the buckling amplitudes in the np-Au remains large enough to cause cracking at the peaks of the buckles, leading to microscopic cracks inside the islands surrounded by the macroscopic cracks.

Finally, the negligible changes in the np-Au ligament and pore widths with the variation in PDMS thickness suggests that there is not a large difference in tensile stress (at least not large enough to influence silver dissolution or gold surface diffusion) in np-Au films on varying PDMS thicknesses.

## 5. Conclusions

The morphology and topography evolution in np-Au thin films at different length scales on rigid glass substrates was compared to that on compliant PDMS substrates anchored to glass supports. In addition, the variation in the film morphology with the change in PDMS thickness (0.25–3.18 mm) was investigated. There was no crack formation in the as-deposited films on glass, but cracking occurred at all length scales after dealloying np-Au on glass. The density of the larger macroscopic cracks for np-Au/glass was on par with that of the films on the thickest PDMS, though they occurred at a slightly smaller length scale due to different underlying mechanisms. The average crack area and crack coverage percentage of the microscopic cracks for np-Au/glass films, however, were more than two-fold higher than the maximum values for any np-Au/PDMS film. The median ligament and median pore widths in np-Au/glass were also markedly higher compared to those in np-Au/PDMS, possibly due to higher tensile stress experienced by the films on the stiffer glass substrate.

The cracking pattern evolved from rigid glass to compliant PDMS, with the extent of cracking changing with the variation in PDMS thickness. Unlike the macroscopic cracks in np-Au/glass, those in np-Au/PDMS formed discrete islands with the island widths decreasing with increasing PDMS thickness, indicating an increase in crack density. The microscopic cracks in np-Au/PDMS showed a similar trend as the macroscopic cracks, with the extent of cracking increasing with PDMS thickness. The microscopic cracks were absent in the films on thinner PDMS substrates, but on thicker PDMS substrates the average crack area and crack coverage percentage increased by several fold. However, the median ligament and median pore widths did not exhibit any significant variation with the change in PDMS thickness.

In summary, by changing the thickness of a compliant substrate and hence modulating its effective elastic modulus at the substrate-thin film interface, the crack architecture across different length scales could be engineered for np-Au thin films. It would likely be possible to fabricate nearly crack-free np-Au films on PDMS by changing the deposition conditions (e.g., using cryogenic sputtering, or pre-straining the PDMS before sputtering) to suppress the post-deposition macroscopic cracking in AuAg and by choosing the optimum thickness of PDMS (~ 0.50 mm) to eliminate the microscopic cracks. Thus, the findings here are expected to inform the design of np-Au functional coatings on compliant substrates for a variety of applications, including wearable sensors.

## Declaration of interests

The authors declare that they have no known competing financial interests or personal relationships that could have appeared to influence the work reported in this paper.

## Data Availability Statement

The datasets generated and/or analyzed during the current study are available from the corresponding author upon reasonable request.

## Acknowledgements

The authors gratefully acknowledge the support from the National Science Foundation via DMR-2003849. Part of this study was carried out at the UC Davis Center for Nano and Micro Manufacturing (CNM2). We thank Prof. Seung Sae Hong's research group for their assistance with the AFM characterization of np-Au thin film topography on different substrates.

**Supporting Information**

# Influence of mechanical compliance of the substrate on the morphology of nanoporous gold thin films

Sadi Shahriar[a], Kavya Somayajula[b], Conner Winkeljohn[a], Jeremy Mason[a], Erkin Seker[c,*]

[a]Department of Materials Science and Engineering, University of California - Davis, Davis, CA 95616, USA

[b]Department of Mechanical and Aerospace Engineering, University of California - Davis, Davis, CA 95616, USA

[c]Department of Electrical and Computer Engineering, University of California - Davis, Davis, CA 95616, USA

## 1. SEM image analysis for characterizing the morphological features

*1.1. Analysis of the discrete islands bound by macroscopic cracks*

Edge-detection and gain extract in GIMP were applied to 8-bit grayscale images to accentuate the cracks against the background. The Sauvola auto local threshold method was applied to the GIMP-processed image to obtain binary segmented images separating the cracks from the background, with the cracks and the areas inside the cracks given pixel values of 0 and 255, respectively. An erosion process was applied to the binary images to close the boundaries of the discrete islands. Finally, the elliptical outlines of the closed regions were obtained, and the major and minor axes lengths of the ellipses were measured in Fiji. Figure SI 1 shows representative images of these processing steps.

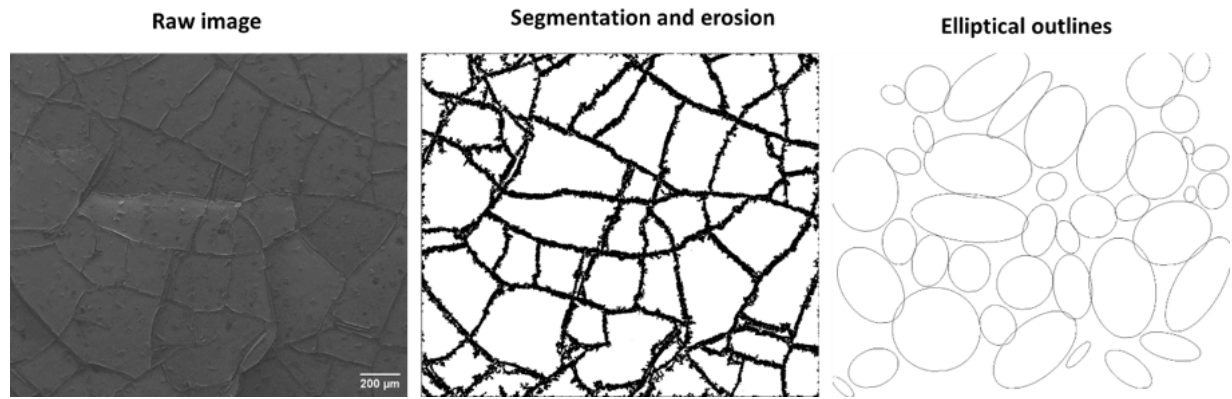

**Figure SI 1**. The process of obtaining elliptical outlines of the discrete islands bound by macroscopic cracks.

*1.2. Analysis of the microscopic cracks*

8-bit images were subjected to trainable Weka segmentation in Fiji to obtain binary images such that the cracks (pixel value of 255) from the background (pixel value of 0). A size threshold of 30 pixels was then applied to filter out the smaller specks that were not cracks. The area of each crack was computed in Fiji to obtain the average crack area in an image. The total crack area in an image was divided by the image area to calculate the crack coverage percentage. Figure SI 2 shows a raw SEM image and the corresponding segmented image showing the cracks in black against a white background.

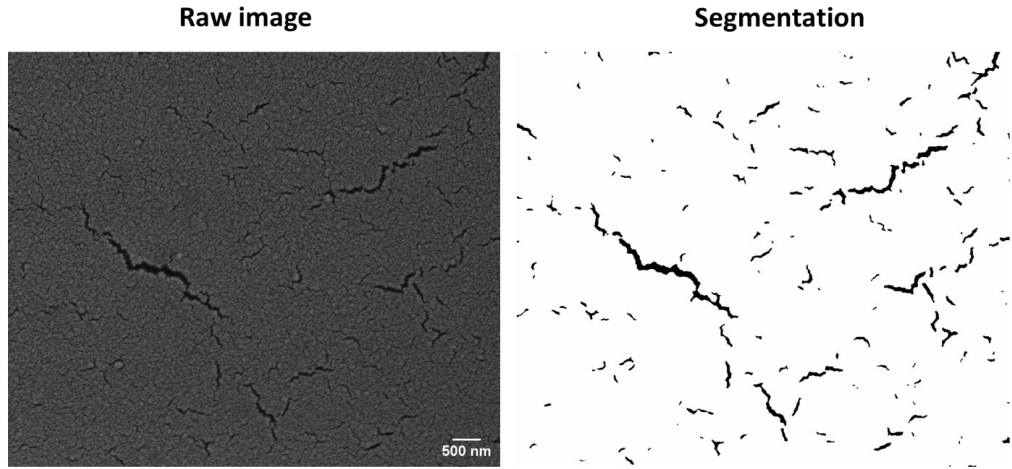

**Figure SI 2**. Binary image obtained from a raw image of the microscopic cracks through Weka segmentation.

*1.3. Analysis of the ligament and pore widths*

8-bit images were subjected to Otsu local auto threshold to turn the images into binary images where pores turn into pure black (0) and the ligaments turn into pure white (255) pixels. The segmentation was followed by the application of a median filter to smoothen the edges of the morphological features. The resulting binary image was considered as a matrix of pixel values consisting of 0 and 255. A chord length distribution (CLD) approach was used to measure the ligament and pore widths where a chord is a line segment with each of its interior points located in one of the two phases (ligament or pore) and the end points touching the interfaces with the other phase. To find the CLD, we used a custom MATLAB script which scans pixel-by-pixel along each row and column of the matrix and records whenever it encounters the pixel value of the desired phase. For example, after the code starts running, each time the algorithm detects a pixel value of 0, it records it as one of end points of a pore chord and stops assigning pixels to that chord after a value of 255 is encountered. This identifies the chord of a specific length corresponding to a pore and after scanning along every row, all the horizontally-oriented pore chords were identified. Similarly, scanning along the columns provided the ligament and pore widths along the vertical axis. Figure SI 3 shows the binary image obtained by thresholding a raw image.

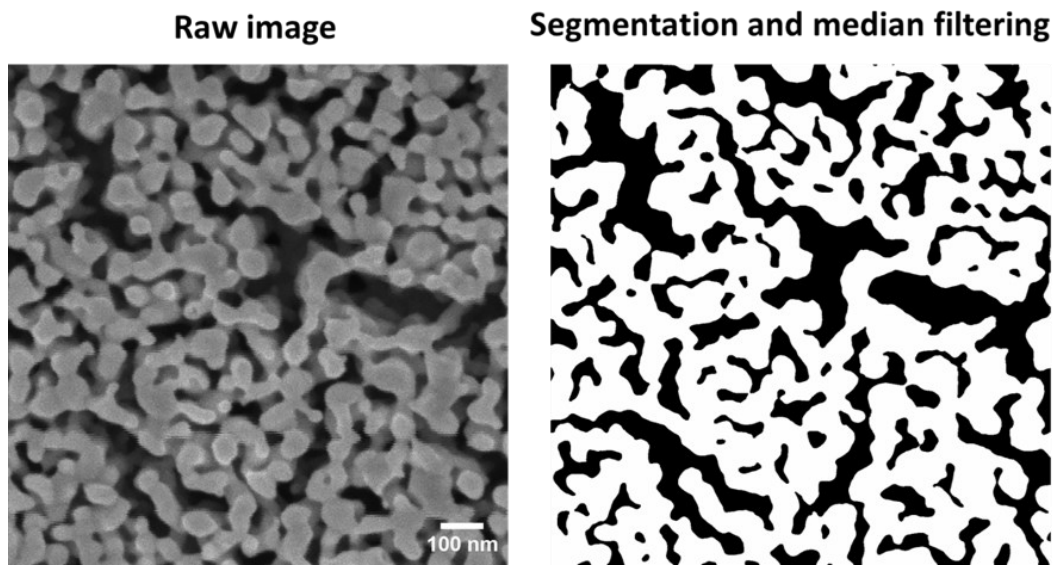

**Figure SI 3**. Binary image of ligaments and pores showing the ligaments in white and the pores in black obtained by thresholding and median filtering the raw image.

*1.4. Analysis of the macroscopic cracks in np-Au/glass*

The macroscopic cracks on glass do not form discrete islands and hence, cannot form elliptical outlines. Therefore, the SEM images for np-Au/glass were made binary to segment the cracks (Figure SI 4) and the MATLAB script for ligament and pore width extraction was applied to the binary images to compute the distances between neighboring cracks. However, distributions of distances found by the two methods do not match exactly since the definitions of crack-to-crack distance are different for the two methods. To make the crack-to-crack distances in np-Au/glass comparable with those in np-Au/PDMS computed by the elliptical outline method, the distances on glass were multiplied by a factor of 1.53. This factor was found by applying the MATLAB script to images of np-Au on 3.18 mm-thick PDMS to measure the crack-to-crack distances. Then, the ratio of the median distances by the elliptical method (~153 µm) to that obtained by the MATLAB code (~100 µm) was used as the upscaling factor for the distances on glass. The rationale for choosing the 3.18 mm-thick PDMS substrates for calculating the upscaling factor is that the density of cracks in the films supported by them is visually similar to those in the glass-supported films.

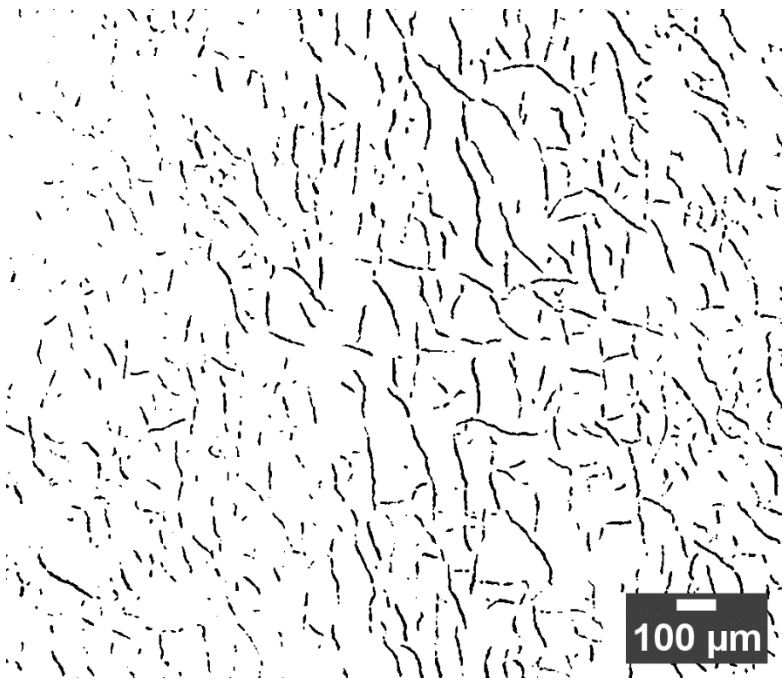

**Figure SI 4**. Binary image showing the segmentation of the macroscopic cracks in np-Au/glass at 150X magnification.

## 2. Image-to-image and sample-to-sample comparison

Three images from each sample (total two samples) at different locations for each type of film-substrate combination were analyzed. Figure SI 5 to SI 7 show the comparisons among the images from each sample where 'Im' denotes image. Figure SI 8 shows the comparison between the samples where 'S' denotes sample.

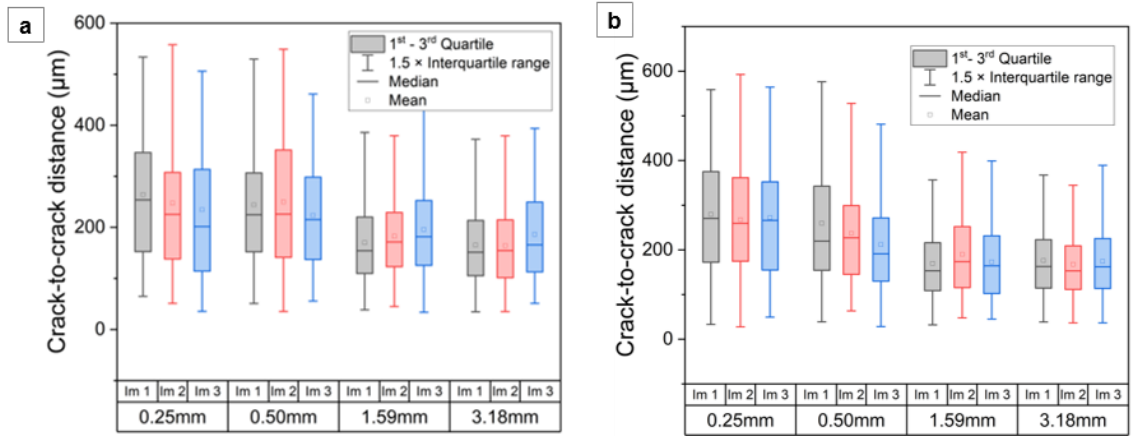

**Figure SI 5.** Macroscopic crack-to-crack distance comparison across images at different locations of (a) sample 1, and (b) sample 2.

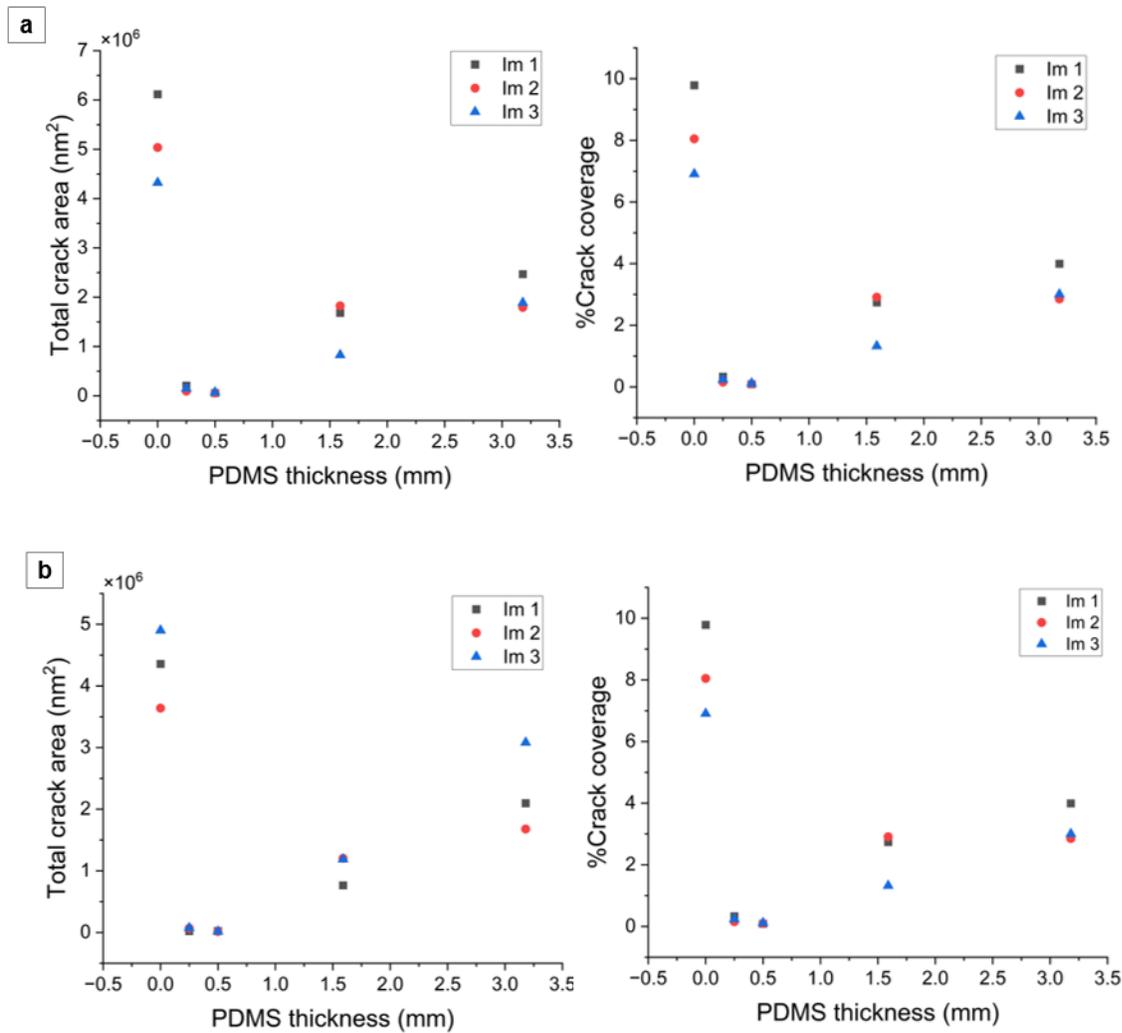

**Figure SI 6.** Total microscopic crack area and crack coverage percentage across images at different locations for (a) sample 1, and (b) sample 2.

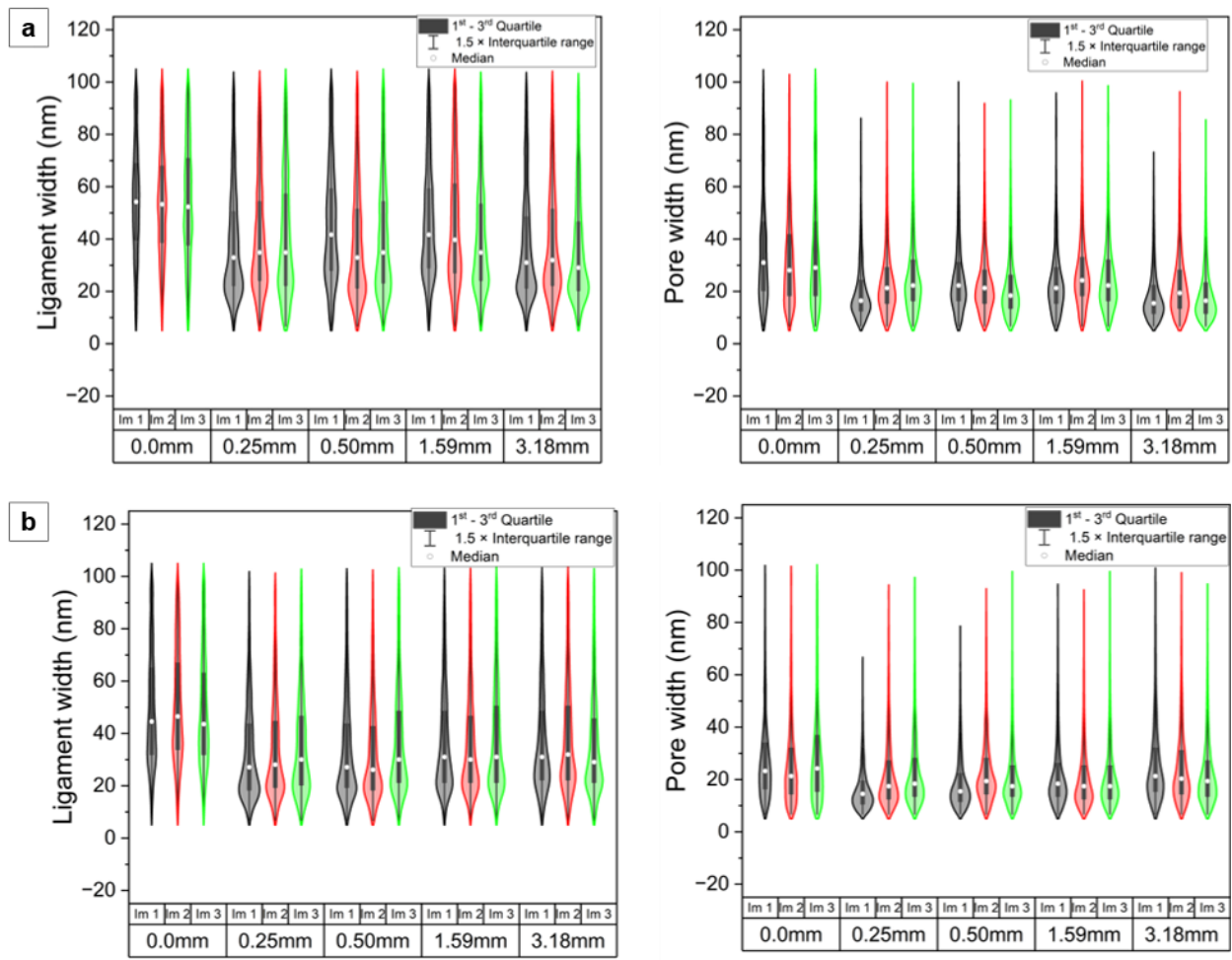

**Figure SI 7**. Ligament and pore widths distribution across images at different locations for (a) sample 1, and (b) sample 2.

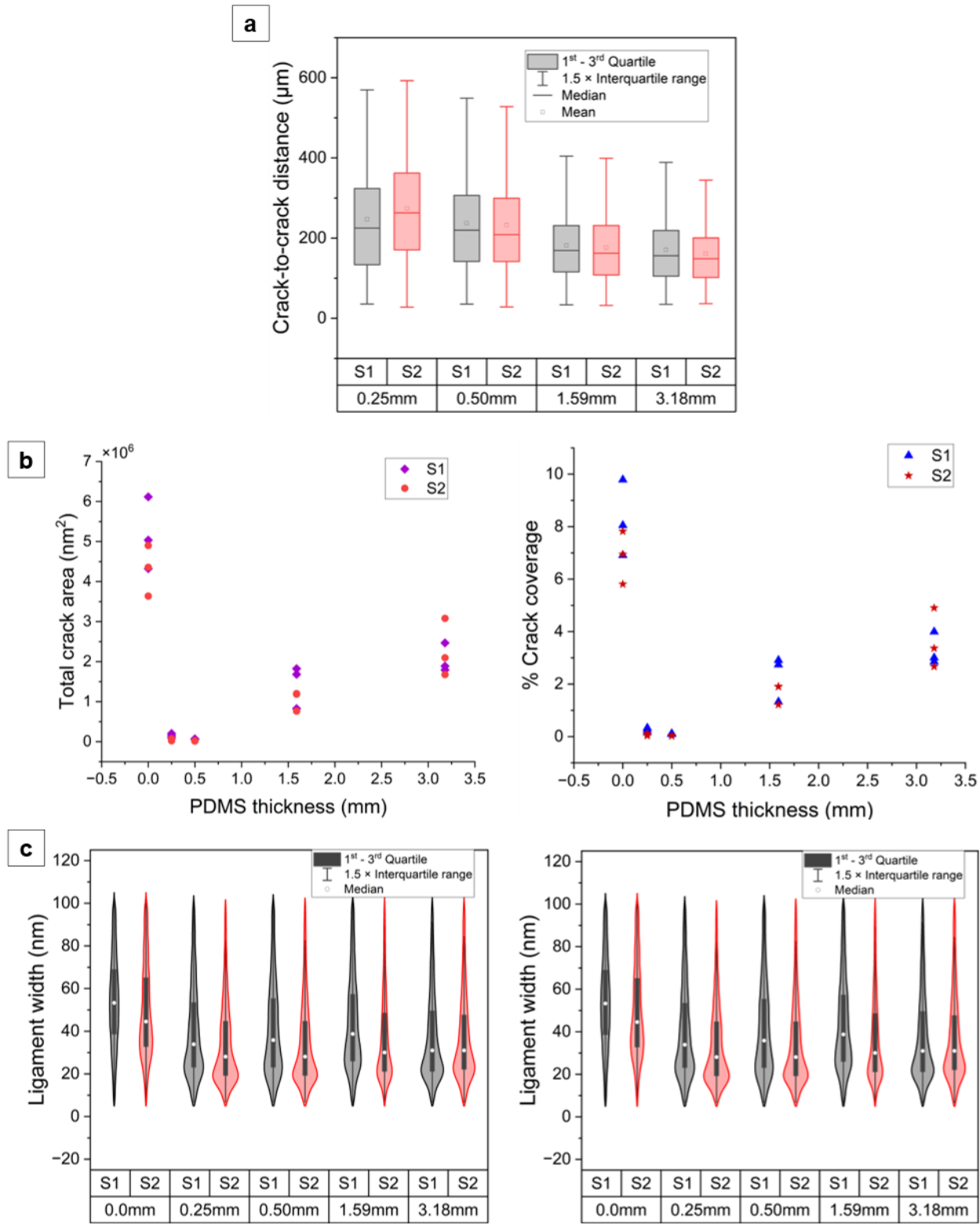

**Figure SI 8**. (a) Macroscopic crack-to-crack distance, (b) Total crack area and crack coverage percentage, and (c) ligament and pore width comparison between sample 1 and 2.

## 3. Additional experiments with thinner glass substrates and higher maximum PDMS thickness

We performed an additional set of experiments where we performed 5 minutes of dealloying at room temperature followed by 15 minutes of dealloying at 55°C for the AuAg films deposited on PDMS substrates of thicknesses 0.25, 3, and 5 mm attached to 0.20 mm-thick glass coverslips. The SEM images of the dealloying-induced microscopic cracks and plots for the distribution of crack widths are shown in Figure SI 9a and Figure SI 9b, respectively.

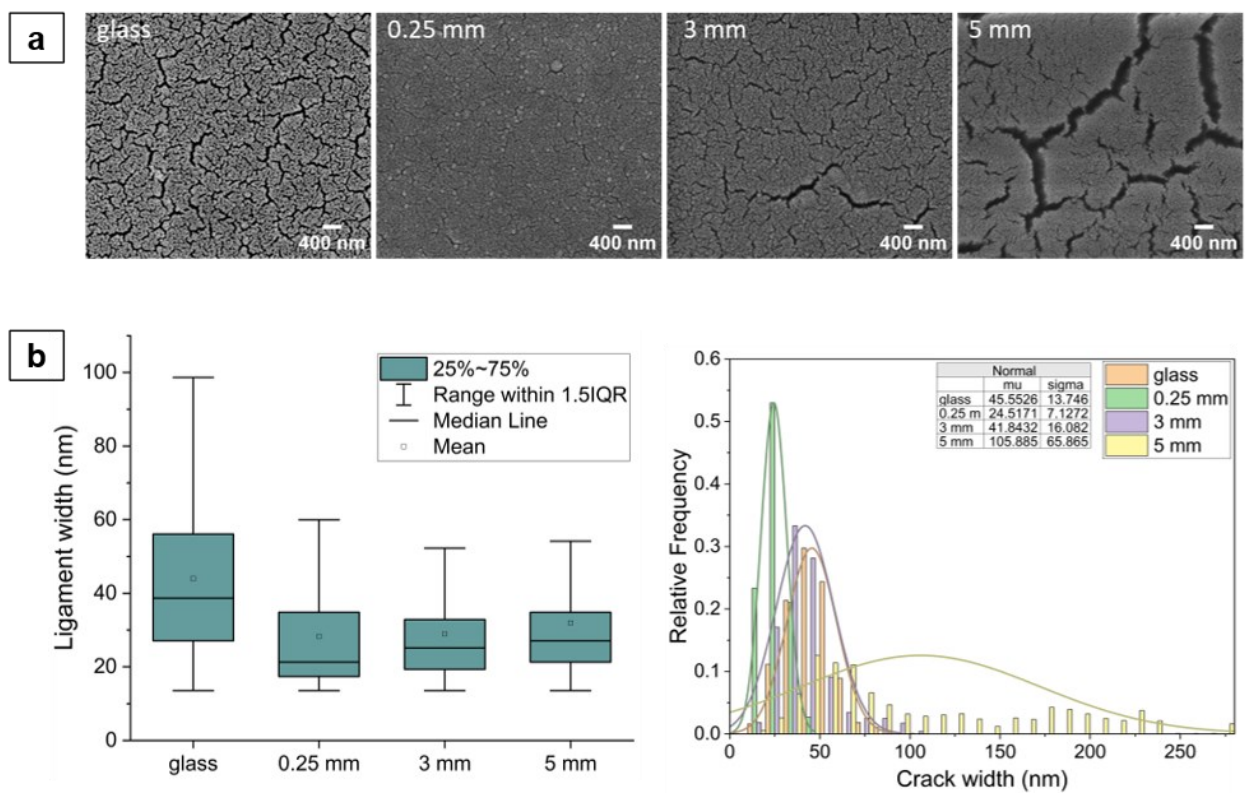

**Figure SI 9**. (a) SEM images of the microscopic cracks in np-Au/glass and np-Au/PDMS thin films, and (b) plots depicting the width distribution of the microscopic cracks. The red lines in the bar plot correspond to the median values and the error bars show the standard deviations.

Figure SI 10 shows tilted SEM images of np-Au films on glass and 1.59 mm-thick PDMS from the main set of experiments showing the differences in the widening of the cracks.

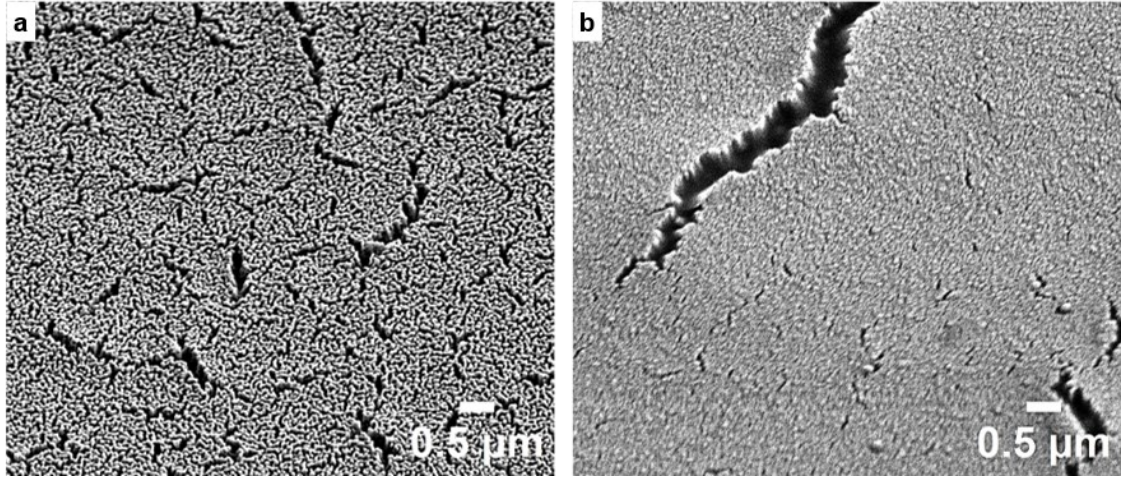

**Figure SI 10**. SEM of microscopic cracks tilted at 45° in (a) np-Au/glass film, and (b) np-Au/1.59 mm PDMS film showing the differences in crack opening in both cases.

### 4. Finite-element simulation methodology

Simulations were performed to estimate a) the effective elastic modulus of PDMS of variable thickness anchored to a glass substrate, b) the dependence of the strain energy of the thin film layers and the PDMS substrates after deposition and dealloying on the PDMS thickness, and c) average horizontal and vertical edge deformation magnitude at the PDMS-metal stack interface as a function of PDMS thickness. The simulations were performed using COMSOL Multiphysics software, and the experimentally-found residual stress value of ~100 MPa was used to estimate the thermal strain experienced by the AuAg, Au, and Cr films following deposition. The corresponding thermal strain was translated into a temperature difference using the coefficient of thermal expansion of the film and fed into the software as the thermal contraction condition. The simulations involved AuAg/Au/Cr layers for the post-deposition and np-Au/Au/Cr layers for the post-dealloying conditions with the PDMS substrates fixed at the bottom to mimic being anchored to rigid glass slides. For the first set of simulations the PDMS thicknesses were 0.01, 0.10, 0.15, 0.25, 0.50, 1.59, 3.18 and 5 mm, and for the second set of simulations were 0.05, 0.10, 0.15, 0.25, 0.50, 0.75, 1, 1.59, 3.18, and 5 mm. The first set of simulations was in 2D, whereas the second set was in 3D. The thickness of the AuAg and the np-Au films in the simulations were 600 nm and 500 nm, respectively. The finite element mesh for the AuAg/Au/Cr/PDMS and np-Au/Au/Cr/PDMS stacks used in the second set of simulations is shown in Figure SI 11.

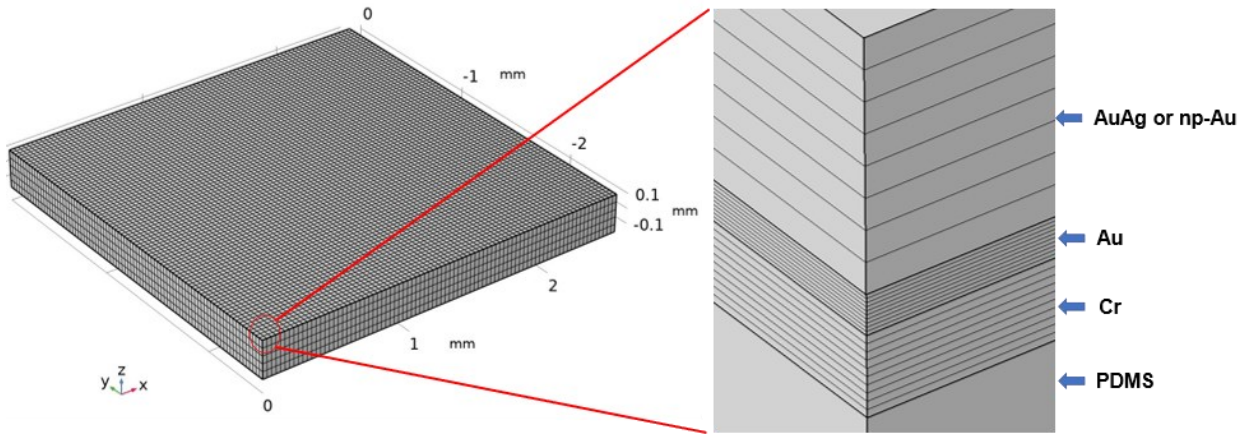

**Figure SI 11.** Finite element mesh for the post-deposition and post-dealloying film/substrate stack.

To estimate the effective surface modulus, two points on the PDMS top surface were displaced axially in the x-direction per the nominal strain values, and the effective stiffness was estimated from the force-displacement plots. The initial separation distance between two such points on the PDMS was fixed at 200 µm. To estimate the variation in strain energy upon dealloying, the thermal strains in the AuAg, Au, and Cr layers are applied as a pre-strain in the dealloying simulations.